# REDS: Random Ensemble Deep Spatial prediction

Ranadeep Daw*[1] | Christopher K. Wikle[2]

[1]Department of Statistics, University of Missouri, Missouri, United States

[2]Department of Statistics, University of Missouri, Missouri, United States

**Correspondence**
*Ranadeep Daw
Email: ranadeepdaw@mail.missouri.com

**Present Address**
Columbia, Missouri, United States.

**Abstract**

There has been a great deal of recent interest in the development of spatial prediction algorithms for very large datasets and/or prediction domains. These methods have primarily been developed in the spatial statistics community, but there has been growing interest in the machine learning community for such methods, primarily driven by the success of deep Gaussian process regression approaches and deep convolutional neural networks. These methods are often computationally expensive to train and implement and consequently, there has been a resurgence of interest in random projections and deep learning models based on random weights – so called reservoir computing methods. Here, we combine several of these ideas to develop the Random Ensemble Deep Spatial (REDS) approach to predict spatial data. The procedure uses random Fourier features as inputs to an extreme learning machine (a deep neural model with random weights), and with calibrated ensembles of outputs from this model based on different random weights, it provides a simple uncertainty quantification. The REDS method is demonstrated on simulated data and on a classic large satellite data set.

## 1 | INTRODUCTION

There is a famous AI hacker koan on the subject of randomness in neural models, reproduced here from the abstract of Rahimi and Recht (2008):

> In the days when Sussman was a novice, Minsky once came to him as he sat hacking at the PDP-6. "What are you doing?" asked Minsky. "I am training a randomly wired neural net to play tic-tac-toe," Sussman replied. "Why is the net wired randomly?" asked Minsky. Sussman replied, "I do not want it to have any preconceptions of how to play". Minsky then shut his eyes. "Why do you close your eyes?" Sussman asked his teacher. "So that the room will be empty," replied Minsky. At that moment, Sussman was enlightened.

One take-home message from this koan is that just as a room is not empty simply because you close your eyes, random networks do have "preconceptions" (useful connections) induced by the randomness, they just are not the ones that we might have learned directly from the data. This helps explain the fact that there has been a great deal of research and development of random network models in the last 20 years or so, and that the randomness presents plausible networks that can be implemented without the significant overhead associated with more standard backpropagation optimization methods. Building on these frameworks, we propose such an algorithm for efficient spatial prediction that uses randomly generated hierarchical (or deep) features, requires minimal training, and uses calibrated ensembles to provide uncertainty quantification (UQ) for the predictions.

Spatial data are often considered to be a realization from a spatially-dependent stochastic process at point locations or fixed areas (see Banerjee, Carlin, & Gelfand 2014; Cressie 1993; Cressie & Wikle 2015; Wikle, Zammit-Mangion, & Cressie 2019, for overview of classical spatial statistics). When such data are reasonably represented by Gaussian processes (GPs) or Markov random fields (MRFs), the primary implementation focus is on the covariance or precision matrices. Yet, it can be challenging to learn these since we often have only one noisy realization from the underlying dependent process, which prohibits the direct estimation of unstructured covariance or precision matrices when working with finite collections of data. Thus, further assumptions such as the specification of a parameterized covariance function, stationarity, and isotropy are often used along with the distributional assumption underlying the data (e.g., GPs or Gaussian MRFs). The computational challenges associated with these methods in the presence of large data sets and prediction domains has led to vigorous research on simplified models and advanced computational approaches to implement spatial prediction algorithms (e.g., see Heaton et al. 2019, for discussion and comparison of several such approaches). Recently, there has been a corresponding interest in the development of machine learning methods that can provide alternative fast, easy-to-implement, and explainable methods to analyze such data without specifying or estimating a covariance model (e.g., Kirkwood, Economou, & Pugeault 2020).

Deep learning (e.g., Goodfellow, Bengio, & Courville 2016) has become very popular over the last decade in nearly every area of data analysis, but particularly with dependent data (e.g., images and sequences). The development of useful tools such as backpropagation, automatic differentiation, stochastic optimization, and GPU computing has allowed researchers and practitioners to extract complex features from dependent data. Deep learning has been applied to the analysis of spatial data in various ways such as using locations or basis expansions of locations as features (Chen, Li, Reich, & Sun 2020; Cracknell & Reading 2014) or in terms of convolutional neural networks (CNNs) (see the recent review of deep learning for spatial and spatio-temporal data in Wikle & Zammit-Mangion 2022). CNNs (Fischer 2006) represent arguably one of the most successful deep learning tool for spatial data, but they are focused typically on using spatial image data as predictors for regression or classification rather than spatial prediction, which is more common in statistics. Deep models are typically over-parameterized and hence, require a great deal of training data and regularization to be successful. As such, they can be challenging to fit and are almost always quite computationally demanding (although this can be mitigated to some extent by pretraining networks on huge image databases). This has motivated researchers to find alternate algorithms that mimic existing deep models while bypassing the expensive training required to learn the weights that connect hidden units. Perhaps the most successful approaches fall under the umbrella of *reservoir computing*, which are methods that select random weights at various layers from some reservoir distribution. Examples of reservoir methods include extreme learning machines (ELMs) (Huang, Zhu, & Siew 2006), echo state networks (ESNs) (Jaeger 2001), and liquid state machines (Maass 2011).

Our goal in this manuscript is to demonstrate that a similar deep random parameter-based model can be used for the prediction of continuous spatial data. Note that classical CNN models require gridded data and often need additional layers to convert such data to a continuous domain problem. Our approach is not limited by a gridded data requirement. However, reservoir approaches face two criticisms in general. First, they can be somewhat "hit and miss" in the sense that some randomly generated networks do not work well by chance. Second, as with classical deep learning models, these reservoir methods do not provide a model-based estimate of the uncertainty associated with predictions. Model-based prediction uncertainty quantification (UQ) is fundamental to statistical spatial prediction procedures. To address the first concern, we have developed an approach to retain the "good" networks from our ensemble of potential networks by using a greedy cross-validation procedure that considers small random subsets of the training data (i.e., minibatches) in every iteration to fit the model, and then examines the fitted model on the remaining data. To address the second concern, we then use these "good" networks in a calibration-based UQ framework. We call our procedure the *REDS* (Random Ensemble Deep Spatial) approach for spatial prediction and uncertainty quantification.

The paper is organized as follows. Section 2 contains relevant background of models and algorithms that have motivated the REDS procedure. Section 3 describes the REDS algorithms and associated methodology. Section 4 shows the simulation and application results, and Section 5 provides a brief conclusion and discussion of future work.

## 2 | BACKGROUND

Randomization has been a fundamental component of statistical modeling since the contributions of R.A. Fisher and the design of experiments (Fisher 1992). Some examples of notable random learning algorithms in statistics include random forests (Ho 1995), the bootstrap (Efron 1992), bagging (Breiman 1996), random projection (Bingham & Mannila 2001; Dasgupta 2013), neural networks (Schmidt, Kraaijveld, Duin, et al. 1992; Suganthan & Katuwal 2021), reservoir computing (Schrauwen, Verstraeten, & Van Campenhout 2007), random Fourier features (Rahimi & Recht 2007), random kitchen sinks (Rahimi & Recht 2008), random vector functional link networks (Pao, Park, & Sobajic 1994), SVD (Halko, Martinsson, & Tropp 2011), random augmentation (Cubuk, Zoph, Shlens, & Le 2020; Zhong, Zheng, Kang, Li, & Yang 2020), coresets (Assadi & Khanna 2017), kernel learning (Xie, Liu, Wang, & Huang 2019), lazy training via neural tangent kernels (Jacot, Gabriel, & Hongler 2018), to name a few. Some of the randomization algorithms (such as reservoir computing) are used to ease the computation cost of their deterministic counterparts, and in some cases are improvements over their deterministic analogs (e.g., random forests, bagging, bootstrap). In general, the simplicity and the computational efficiency of randomized algorithms has attracted researchers and practitioners to these methods.

Classical spatial data analysis methods such as kriging (or GP regression) can be computationally challenging in high dimensions. For example, spatial data at $n$ locations have an $n \times n$ covariance matrix that must be inverted with complexity $\mathcal{O}(n^3)$, which is problematic when the number of observations is over several thousand. This has necessitated the development of scalable methods for spatial prediction. Some scalable approaches include pseudo-likelihood methods (Besag 1975), (sparse) precision matrices (Leroux, Lei, & Breslow 2000), graphical models (Kim & Nelson 1999), variational Bayes, predictive processes (Banerjee, Gelfand, Finley, & Sang 2008), to name a few. The precision matrix based approximations usually assume conditional independence in the data given a neighborhood, which leads to sparse precision matrices. This has led to the development of nearest neighbor based GPs and Vecchia approximations (e.g., (Datta, Banerjee, Finley, & Gelfand 2016; Katzfuss, Guinness, Gong, & Zilber 2020; Vecchia 1988). One can also introduce sparsity in the covariance matrix directly through tapering (Furrer, Genton, & Nychka 2006; Kaufman, Schervish, & Nychka 2008).

An alternative to the neighbor-based methods for spatial and spatio-temporal GPs is to use kernel or basis function approximations (e.g., Higdon 1998; Wikle & Cressie 1999). There have been many implementations of such approaches, with a variety of basis functions and with some methods being low rank, full rank, or overcomplete (see Cressie, Sainsbury-Dale, & Zammit-Mangion 2021; Hefley et al. 2017; Wikle 2010, for overviews). There have been recent developments in low-rank basis expansions in the machine learning community. For example, *random Fourier features (RFFs)* (Rahimi & Recht 2007) provide extra flexibility in modelling spatial data (see Hensman, Durrande, Solin, et al. 2017; Miller & Reich 2022; Milton, Coupland, Giorgi, & Bhatt 2019). RFFs have also used in a deep feature expansion context (e.g., Cutajar, Bonilla, Michiardi, & Filippone 2017; Xie et al. 2019). Deep RFFs derive the features by using a set of hierarchical RFF transformation of the features. As discussed below, REDS considers a similar deep network and provides more flexibility in generating features by using a combination of RFFs and other randomly generated features. REDS first transforms the spatial coordinates into a latent space of the RFFs. Since these shallow features are often not expressive enough to accommodate the dependence structure in high volume spatial processes, we further look to generate more complex features. Our approach is motivated by Chen et al. (2020), where a combination of basis expansions and a deep feed-forward neural network was used. However, we seek to find a more computation-friendly approach that is free of the training burden of the deep neural network. Thus, we replace the deep networks with Extreme Learning Machines (Huang et al. 2006), which use randomly-weighted reservoirs to project the inputs on to a more complex feature space. Since ELMs replicate the deep networks in a random fashion, our approach generates more flexible features than the deep RFFs. In other words, our method is a computationally cheaper alternative of the Deepkriging algorithm that is suitable for large datasets.

All viable spatial prediction methodologies must include uncertainties associated with the predictions. This arises naturally in GP-based approaches. Although UQ is not produced from traditional deep neural models, there have been numerous approaches that accommodate UQ in these models, and it is an expanding area of research (see the review in Abdar et al. 2021). For example, perhaps the first approach for UQ in neural networks was mixture density networks (Bishop 1994). These consider a mixture of Gaussians to arbitrarily approximate the residual density it uses the deep network to predict the parameters of the mixture distribution. Another more recent example is "Monte Carlo dropout", which generates a set of dropout layers (Srivastava, Hinton, Krizhevsky, Sutskever, & Salakhutdinov 2014) that randomly remove connections in the hidden layer and produces the corresponding set of predictions using those layers. The set of predictions are then used to construct the prediction bounds. Other methods include Bayesian neural networks (McDermott & Wikle 2019a; Wang & Yeung 2016), "Bayes by Backprop" (Blundell, Cornebise, Kavukcuoglu, & Wierstra 2015), and deep GPs (Damianou & Lawrence 2013). Simpler alternatives for UQ have been developed based on discrete density estimation as in Li, Bondell, and Reich (2019) (which is the approach used in DeepKriging). A cheaper alternative is to consider ensembles, which are particularly useful in random network approaches such as reservoir computing (e.g., McDermott & Wikle 2017 2019b). Such ensemble approaches should be calibrated to achieve a particular coverage probability (e.g., Bonas & Castruccio 2021). In the REDS algorithm, we consider a very simple UQ calibration approach motivated by the procedure outlined in Sørbye and Rue (2014), which was originally developed for prior elicitation in MRF spatial models.

## 3 | REDS METHODOLOGY

In this section we describe the methodology for the REDS prediction approach, including point prediction (Section 3.1) and associated uncertainty quantification (Section3.2).

### 3.1 | Prediction Methodology

Consider spatial locations $S = \{s_1, \ldots, s_n\}$ with corresponding data represented by $Y = \{Y_1, \ldots Y_n\}$. Our only assumption here is that we are dealing with a spatially-dependent random variable $Y_i$ that has a continuous response; we make no assumptions as to the nature of the spatial dependence (e.g., functional form, isotropy, stationarity). The goal here is to provide predictions at some locations of interest $\tilde{S} = \{\tilde{s}_1, \ldots, \tilde{s}_{\tilde{n}}\}$ where the realization of $Y$ may not be available. Along with this prediction task, we also try to estimate the uncertainty of our predictions. To better illustrate our approach, we first present the REDS point prediction algorithm (Algorithm 1), and then describe the methodological motivation corresponding to each step.

---

**Algorithm 1** REDS: Generating an ensemble of predictions

---

1: **Input** Training locations $S = \{s_1, \ldots, s_n\}$, training data $Y = \{Y_1, \ldots Y_n\}$, minibatch size $M$, test locations $\tilde{S} = \{\tilde{s}_1, \ldots, \tilde{s}_{\tilde{n}}\}$
2: Draw $\boldsymbol{\omega}_1, \ldots, \boldsymbol{\omega}_J \sim p(\boldsymbol{\omega})$.
3: $h^{(0)}(s) = \left( \cos(\boldsymbol{\omega}_1^T s), \sin(\boldsymbol{\omega}_1^T s), \ldots, \cos(\boldsymbol{\omega}_J^T s), \sin(\boldsymbol{\omega}_J^T s) \right)^T$.
4: **for** $j = 1 : N$ **do**
5:     Draw number of hidden layers $K_j \sim \pi_K$, hidden layer width $d_j \sim \pi_d$, penalty parameter $\lambda_j \sim \pi_\lambda$.
6:     **if** $K_j > 0$ **then**
7:         Draw $\boldsymbol{V}^{(\ell)}$ with elements $v_{kl}^{(\ell)} \overset{iid}{\sim} q(v)$ for $l = 1, \ldots, K_j - 1$.
8:         $h^{(\ell+1)}(s) = \max \left( \boldsymbol{V}^{(\ell)} h^{(\ell)}(s), 0 \right)$ for $l = 1, \ldots, K_j - 1$.
9:     **end if**
10:     Select a random subset $\check{S}$ from $S$ of size $M$.
11:     Fit the model: $Y_{\check{s}} = p\left( h^{(K)}(\check{s}); \lambda_j \right)$ using $\{Y_{\check{s}} : \check{s} \in \check{S}\}$.
12:     Predict $\hat{Y}_{\check{s}}^{(j)} = \hat{p}\left( h^{(K)}(\check{s}); \lambda_j \right)$ for $\check{s} \in \check{S}$ (similarly, predict $\hat{Y}_s^{(j)}$ for $s \in S \cup \tilde{S}$).
13:     Compute the training RMSEs as $\epsilon_j = \sqrt{\frac{1}{n} \sum_{s \in S} \left( \hat{Y}_s^{(j)} - Y_s \right)^2}$.
14: **end for**
15: Order the training RMSEs as $\epsilon_{(1)} \leq \ldots \epsilon_{(R)} \leq \ldots \leq \epsilon_{(N)}$.
16: Define $\mathcal{J} \subseteq \{1, \ldots, N\}$ as the set of good networks corresponding to $\epsilon_{(1)}, \ldots, \epsilon_{(R)}$.
17: Ensemble Prediction: $\hat{Y}_{en,s} = \text{median}(\hat{Y}_s^{(j)}) : j \in \mathcal{J}$ for $s \in S \cup \tilde{S}$.
18: Ensemble Uncertainty: $\hat{\sigma}_{en,s} = \text{IQR}(\hat{Y}_s^{(j)}) : j \in \mathcal{J}$ for $s \in S \cup \tilde{S}$.
19: **Output** $\hat{Y}_{en,s}$ and $\hat{\sigma}_{en,s}$ for $s \in S \cup \tilde{S}$.

---

Step 2 and 3 of Algorithm 1 transforms the input to a random basis layer using a set of random frequencies from distribution $p(\omega)$. The output features ($\{\cos(\boldsymbol{\omega}_j^T s), \sin(\boldsymbol{\omega}_j^T s)\}$) are known as random Fourier features (RFFs), as introduced by Rahimi and Recht (2007); see Liu, Huang, Chen, and Suykens (2020) for a review of different RFF approaches. RFFs are motivated by the well-known Bochner's theorem (Rudin 2017), which states that any positive definite and shift invariant kernel $\mathcal{K}(\boldsymbol{x}, \boldsymbol{y}) : S \times S \to \Phi$ can be rewritten as the Fourier transform of a probability density $p(\boldsymbol{\omega})$ as the following

$$\mathcal{K}(\|\boldsymbol{x} - \boldsymbol{y}\|) = \int \exp\{i\boldsymbol{\omega}^T \|\boldsymbol{x} - \boldsymbol{y}\|\} \, p(\boldsymbol{\omega}) \, d\boldsymbol{\omega}. \tag{1}$$

The consequence of the above representation is that it allows us to approximate the kernels using a Monte Carlo evaluation of equation 1 by simulating random frequencies $\Omega = \{\boldsymbol{\omega}_1, \ldots, \boldsymbol{\omega}_r\}$ from $p(\boldsymbol{\omega})$. The kernel $\mathcal{K}$ is then approximated by the following Monte Carlo summation

$$\mathcal{K}(\|\boldsymbol{x} - \boldsymbol{y}\|) \approx \frac{1}{r} \sum_{j=1}^{r} \exp\{i\boldsymbol{\omega}_j^T \|\boldsymbol{x} - \boldsymbol{y}\|\} = \frac{1}{r} \langle \boldsymbol{\phi}(\boldsymbol{x}), \boldsymbol{\phi}(\boldsymbol{y}) \rangle. \tag{2}$$

The last equality in Equation (2) follows from the *kernel trick* (Berlinet & Thomas-Agnan 2011), where $\boldsymbol{\phi}(s)$ are feature vectors corresponding to the data $\boldsymbol{Y}$. In the RFF case, the most common choice of $\boldsymbol{\phi}$ is $\boldsymbol{\phi}(s) = \left(\cos(\boldsymbol{\omega}_1^T s), \sin(\boldsymbol{\omega}_1^T s), \ldots, \cos(\boldsymbol{\omega}_J^T s), \sin(\boldsymbol{\omega}_J^T s)\right)^T$, which is used in the step 3 of Algorithm 1. Hence, step 3 is a fully random basis expansion following a certain family. In our work, we chose the frequency distribution $p(\boldsymbol{\omega})$ in such a way that allows for multi-resolution training, which has often been used in kriging to approximate the associated covariance matrix for a finite sample of observation locations (e.g., Chen et al. 2020; Nychka, Bandyopadhyay, Hammerling, Lindgren, & Sain 2015). We describe our choice of $p(\boldsymbol{\omega})$ in Section 3.3.

Step 5 randomly simulates the hyperparameters of each network, i.e., the number of hidden layers, the hidden layer width, and the lasso regression penalization parameter from distributions described in Section 3.3. Following these choices, step 7 and 8 in Algorithm 1 perform a series of (deep) non-linear stochastic transformations that generates a collection of potential features. This is the same as passing the input through a deep ELM. Recall, a deep ReLU network model takes an input $\boldsymbol{z}$ and performs a series of transformations as in: $\boldsymbol{h}_0 = \boldsymbol{z} \xrightarrow{A_1} \boldsymbol{h}_1 = max(\boldsymbol{A}_1\boldsymbol{h}_0, 0) \xrightarrow{A_2} \ldots \xrightarrow{A_\ell} \boldsymbol{h}_\ell = max(\boldsymbol{A}_\ell \boldsymbol{h}_{\ell-1}, 0) \xrightarrow{A_{\ell+1}} \boldsymbol{A}_{\ell+1}\boldsymbol{h}_\ell = \hat{Y}$. In a classical deep network, all the weight matrices $\boldsymbol{A}_j$ are learned using an optimization algorithm such as stochastic gradient descent. In order to increase the computational efficiency and avoid the variance in the gradient propagating through different layers, ELMs do not estimate the intermediate weight matrices $\boldsymbol{A}_1, \ldots, \boldsymbol{A}_\ell$ and instead generate them from a common distribution such as a Gaussian or uniform distribution. Only the last layer is learned, and can be done so using any simple regularization learning algorithm such as a penalized regression or Bayesian regression (or analogous regularization approaches for non-Gaussian responses). In our case, $\boldsymbol{V}^{(\ell)}$ are the random weight matrices, which are simulated from a distribution $q(v)$. The random projection is then rectified through a ReLU transformation in step 8.

The final training is then done with a randomly selected minibatch (i.e., a random subsample) in steps 10 - 11 using a lasso regression with the penalty parameter $\lambda_j$ chosen in step 5. This avoids overfitting from an overcomplete set of random basis functions and also helps reject the poor performing networks in two ways. First, penalized regression such as ridge, lasso, elastic net, allows shrinkage on the regression coefficients to avoid overfitting. Second, fitting with penalized regression only on a minibatch sample avoids using excess data and hence provides additional protection to avoid overfitting. Minibatch training is often used in deep learning methods such as stochastic gradient descent (Amari 1993). The training root mean square error (RMSE) ($\epsilon_j$) is then computed by calculating the root mean square errors over the full training set in step 13.

Note that after computing the first hidden layer $h^{(0)}$ of the RFFs, the remaining tasks are performed $N$ times. This gives us an ensemble set of $N$ training RMSEs, $\epsilon_{(1)} \leq \ldots \epsilon_{(N)}$, out of which we keep only the best $R$ networks. This allows us to remove the poor performing networks that are sometimes generated in reservoir models. We define $\mathcal{J}$ as the set of indices corresponding to the ensemble members that are retained. The final prediction is taken as the ensemble median (step 17). In addition, for the UQ procedure describe in Section 3.2, we also require a measure of the ensemble uncertainty, for which we used the ensemble inter quartile range (IQR) (step 18). Predictions and IQR values for each spatial location in the training sample, $\boldsymbol{S}$, and for the locations without associated responses, $\bar{\boldsymbol{S}}$, are output from the algorithm (step 19).

## 3.2 | REDS: Uncertainty Quantification

Algorithm 1 produces a set of ensemble median point predictions and IQR estimates for spatial locations of interest, but does not provide a calibrated measure of prediction uncertainty. Thus, we consider a simple approach for UQ that calibrates the ensemble predictions and uncertainties from the output of this algorithm. Specifically, our approach is motivated by the procedure outlined in Sørbye and Rue (2014) that was developed for prior elicitation in MRF spatial models. As outlined in Algorithm 2, the approach uses the set of $R$ ensemble predictions and uncertainties from Algorithm 1 to construct a prediction bound as follows. We define the lower and upper confidence intervals for a given cutoff $v$ as $L_v(.)$ and $U_v(.)$ and the corresponding coverages are denoted by $C_v$. In step 2 and 3 of Algorithm 2, we compute $L_v(s) = \hat{Y}_{en,s} - v\hat{\sigma}_{en,s}$ and $U_v(s) = \hat{Y}_{en,s} + v\hat{\sigma}_{en,s}$ and then compute the coverage $C_v = \frac{1}{n} \sum_{s \in S} \mathbb{1}\left[L_v(s) < Y_s < U_v(s)\right]$ using $s$ from the training dataset, $\boldsymbol{S}$. Then in step 4, $\hat{v}$ is optimized using a simple Brent's bisection method (Brent 1971) to calibrate the optimal cutoff such that $C_{\hat{v}} = \alpha$, which is the target level of the confidence. Thus, this gives the prediction bounds $\left\{ \hat{Y}_{en,s} \pm \hat{v}\hat{\sigma}_{en,s} \right\}$ that contain $100(1-\alpha)\%$ of the observations.

---

**Algorithm 2** REDS: Ensemble Uncertainty Quantification

---

1: **Input** The outputs from algorithm 1, i.e., $\hat{Y}_{en,s}$ and $\hat{\sigma}_{en,s}$ for $s \in S$ (and similar outputs for the locations in $\tilde{S}$), a target Type I confidence level $\alpha$, and an initial cutoff value, $v$.
2: Given $v > 0$, define the confidence bounds as $L_v(s) = \hat{Y}_{en,s} - v\hat{\sigma}_{en,s}$ and $U_v(s) = \hat{Y}_{en,s} + v\hat{\sigma}_{en,s}$.
3: Compute the coverage as $C_v = \frac{1}{n} \sum_{s \in S} \mathbb{1}\left[ L_v(s) < Y_s < U_v(s) \right]$.
4: Compute the optimal calibration cutoff $\hat{v}$ such that $C_{\hat{v}} = 1 - \alpha$.
5: Compute the prediction intervals for each $s$ in the test set $\tilde{S}$ as $\left( \hat{Y}_{en,s} \pm \hat{v}\hat{\sigma}_{en,s} \right)$.

---

## 3.3 | Selection of Hyperparameters

Deep learning algorithms are notorious for having a large number of tuning hyper-parameters. Random learning algorithms are no exception. These tuning parameters often have a significant impact on how good the random layers perform and typically lead to choices that are non-convex, meaning that a linear update of the parameters would not provide a monotonic increase or decrease in the model performances. We discuss the rationale behind our choices of the hyper-parameters, $\theta = \{ J, N, \beta, d, K, \lambda \}$, below.

Like any machine-learning model, REDS requires extensive validation strategy to find the best networks. Our experiments with simulated data suggest that instead of using one good value for each hyperparameter, prediction and coverage results improve when we use different values for hyperparameters across the different minibatches. Therefore, we consider parameters selected randomly from intervals instead of specific values to tune REDS. Such intervals do not need to be mutually exclusive and instead can consider overlapping ranges as shown below. The primary message here is that different minibatches may require different choices of hyperparameters and therefore random selection of hyperparameters improves our results. It is also easier to find a continuous range of values rather than searching for the 'best' points since the RMSEs are often nonlinear in terms of these hyperparameters.

We find that there are some hyperparameters in REDS do not require tuning. The ELM parameter distribution and the RFF distribution are two such examples. The ELM distributions are selected from typical choices (Gaussian/ uniform). Varying the associated hyperparameters showed little to no effect on results during our simulation testing. The RFF distribution is chosen to represent a multi-resolutional basis expansion. This distribution can sometimes change the prediction results results, but our experiments with Gaussian process data suggested that our choice is robust for moderate to large-sized datasets. Therefore, we kept these hyperparameters fixed and investigated the effect of varying other hyperparameters. However, for some applications, it might be beneficial to tune the hyperparameters that we have fixed. The following list describes the choices for hyperparameter selection for those parameters that were not fixed.

i *Number of RFFs ($J$)*: In our examples, we followed the guidance in Chen et al. (2020) for the RFF basis expansion. Layer-0, i.e., the RFF expansion, is specified in a multi-resolution fashion. Specifically, we convert the spatial region into a $[0 \times 1]^2$ square grid and then select three different resolutions of radii, $0.1, 0.2, 0.3$. For each of these resolutions we simulate 500, 300 and 200 random frequencies, respectively, since it is usually the case that the small scale variation needs comparatively more expansion features to generate an accurate representation. This leads to a total of $J = 500 + 300 + 200 = 1000$ basis functions. Since each $\omega_j : j = 1, \ldots, J$ gives two features, $\cos(\omega_j^T s), \sin(\omega_j^T s)$, the dimension of the layer-0 hidden layer $h^{(0)}(s)$ is $2(1000) = 2000$. We use these choices for our simulations and real data example. It should be noted that computation of the cosine and sine functions takes more time compared to the other matrix computations used here and so the generation of RFFs is performed only once.

ii *Distribution of Hidden Layer Elements (q)*: A common choice for $q$ is a Gaussian or uniform distribution. In exploratory studies, we found that the uniform distribution typically performed better than the Gaussian distribution. We also saw that the model predictions do not vary much with the choice of the range of the uniform distribution. Thus, we used a uniform distribution between $-0.1$ and $0.1$ for the choice of $q$.

iii *Distribution of Number of Hidden Layers ($\pi_K$)*: An advantage of the ensemble approach used in REDS is that we can draw important hyper-parameters from a distribution for each ensemble member. This is helpful for the number of hidden layers. Specifically, we select the number of hidden layers $K_j$ corresponding to the $j$-th network randomly from a distribution $\pi_K$. We allow $\pi_K$ to be a discrete uniform distribution over the following ranges: $1-2$, $1-3$, $1-4$. The deeper networks were not particularly helpful in our applications and also increase the computational cost in the large examples.

iv *Distribution of Hidden Layer Width ($\pi_d$)*: First note that, for simplicity, we use the same width for each hidden layer (although, this need not be the case). We experimented with various widths between 500 and 2000 and noticed that the choice of this parameter directly effects the run-time of the algorithm since the computational complexity is proportional to $d_j^2$. In the validation study, the different distributions considered for the hidden width are: $200-500$, $500-1000$, $1000-1200$. In each case, the width is sampled uniformly from the given range.

v *Distribution over penalization parameter ($\pi_\lambda$)*: We experimented with various choices of the lasso penalization parameter. The ranges we considered for $\lambda$ are: $0.001-0.005$, $0.005-0.01$, $0.01-0.05$. Note that we also experimented with ridge and elastic net penalization methods and lasso was slightly better.

vi *Number of ensembles ($N$), "good" ensembles ($R$) and minibatch size ($M$)*: Our experiments showed that although the accuracy results were similar for $N$ of moderate size, the smoothness of the prediction boundary and the coverage values improve with increasing $N$. For small to medium sized datasets, one can fit several thousand ensembles to get results within a minute. For large datasets, it becomes crucial to manage the computational time. For the two large examples ($n$ of the order of $10^5$), we can fit $N = 500$ ensemble networks efficiently.

To choose $R$, we investigated the RMSEs over the training set. We found that most of the networks are quite poor compared to the best available results. Thus, we considered the best 20% of the networks based on RMSE, which left us with $R = 100$ networks on which to base the ensemble prediction and coverage calculations.

Finally, the minibatch size $M$ is kept fixed at $\lceil n/10 \rceil$ (one tenth of the training sample).

## 3.4 | Implementation and Comparison Metrics

We implemented the REDS algorithms using MATLAB (MATLAB 2018) on a 40 core high-performance cluster. The calculations for each ensemble member are implemented in parallel. Note that simulation examples with a few thousand observations can be solved within a few seconds, whereas the two large data examples in Section 4 take around $5-6$ minutes with 500 ensembles. The predictive performance is evaluated using five metrics as described below.

Prediction accuracy is evaluated using two metrics – mean absolute error (MAE) and root mean square error (RMSE). For a set of $n$ test locations $\tilde{S}$ with observed realizations $\{Y_s : s \in \tilde{S}\}$ and predictions $\{\hat{Y}_s : s \in \tilde{S}\}$, these metrics are defined as follows:

$$\text{MAE}(\tilde{S}) = \frac{1}{n} \sum_{s \in \tilde{S}} |Y_s - \hat{Y}_s|,$$

$$\text{RMSE}(\tilde{S}) = \sqrt{\frac{1}{n} \sum_{s \in \tilde{S}} (Y_s - \hat{Y}_s)^2}.$$

We also consider three UQ summary metrics: prediction interval coverage (CO), interval score (IS), and continuous rank probability score (CRPS) (Gneiting & Raftery 2007). Specifically, consider a location $s$ with observation and point prediction $Y_s$ and $\hat{Y}_s$, respectively, and let $\hat{\tau}_s$ denote the standard deviation of our prediction. Also, let the $100(1-\alpha)\%$ prediction interval for this location be denoted by $[L_s, U_s]$. Then, for a particular location, $s$,

$$\text{CO}_s = \mathbb{1}(L_s < Y_s < U_s),$$

$$\text{IS}_s = (U_s - L_s) + \frac{2}{\alpha}\left[(L_s - Y_s)\,\mathbb{1}(Y_s < L_s) + (Y_s - U_s)\,\mathbb{1}(Y_s > U_s)\right],$$

$$\text{CRPS}_s = \hat{\tau}\left[\frac{1}{\sqrt{\pi}} - 2\psi\left(\frac{Y_s - \hat{Y}_s}{\hat{\tau}_s}\right) - \left(\frac{Y_s - \hat{Y}_s}{\hat{\tau}_s}\right)\left(2\Psi\left(\frac{Y_s - \hat{Y}_s}{\hat{\tau}_s}\right) - 1\right)\right].$$

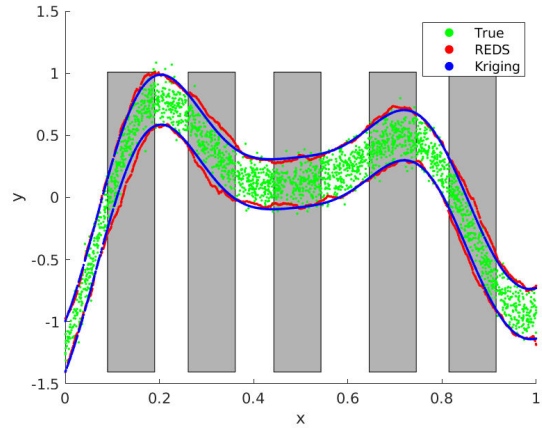

**Figure 1** *One-Dimensional Stationary Gaussian Process Example*: With the best hyperparameters selected as described in Section 3.3, REDS is applied to data simulated from a stationary GP in one dimension. To train REDS, only the data with white background is used and the data with gray background is used as the test set. The green points are the data and the red line shows the prediction interval from a GP regression (kriging) model with the covariance function known. The blue lines are the prediction intervals generated from the REDS algorithm. Note that REDS is providing a prediction bound that is quite similar to that obtained from the kriging model.

where $\psi(.)$ and $\Psi(.)$ are the standard normal pdf and cdf, respectively. Then, given a set of $n$ test locations $\tilde{\boldsymbol{S}}$, the prediction interval coverage, interval score, and the continuous rank probability score for the test set is given by

$$\mathrm{CO}(\tilde{\boldsymbol{S}}) = \frac{1}{n} \sum_{s \in \tilde{S}} \mathrm{CO}_s,$$

$$\mathrm{IS}(\tilde{\boldsymbol{S}}) = \frac{1}{n} \sum_{s \in \tilde{S}} \mathrm{IS}_s,$$

$$\mathrm{CRPS}(\tilde{\boldsymbol{S}}) = \frac{1}{n} \sum_{s \in \tilde{S}} \mathrm{CRPS}_s.$$

# 4 | SIMULATION AND APPLICATION RESULTS

In this section we demonstrate the REDS algorithm on two simulated datasets and real world satellite data. In particular, we use REDS to generate point predictions and uncertainty bounds for these datasets.

## 4.1 | Simulation: One-Dimensional Gaussian Process

Consider simulated data that follow a stationary GP over a univariate domain between 0 and 1 with covariance function given by a Gaussian kernel: $c_{ij} = cov(t_i, t_j) = \exp^{-16(t_i - t_j)^2} + 0.1\delta_{ij}$. This process is simulated at $n = 2500$ locations generated from a uniform distribution between 0 and 1.

We randomly selected intervals of width 0.1 and assigned the data within those intervals to be missing while training the model. We evaluated the performance of REDS at these missing locations for validation. We used the hyperparameter tuning strategy outlined in section 3.3 to find the best fitting hyperparameters. In particular, we used $1 - 4$ hidden layers, $300 - 500$ hidden units, and $\lambda$ in the range of $0.001 - 0.005$. We compared the REDS results to a full GP regression (kriging) implementation assuming that form of the true covariance kernel was known. Table 1 shows the prediction and UQ metric comparisons. The metrics for the REDS predictions are quite similar to those for GP regression, even though the GP regression had the advantage of knowing the true data generating process. Figure 1 shows the comparison between the REDS and GP regression fits and demonstrates that the REDS prediction intervals are quite similar to the GP prediction intervals.

| Metric | MAE | RMSE | CO | CRPS | IS |
|--------|------|------|------|------|------|
| REDS | 0.0785 | 0.0990 | 0.9737 | 0.0566 | 0.5056 |
| Kriging | 0.0768 | 0.0969 | 0.9634 | 0.0546 | 0.4568 |

**Table 1** *One-Dimensional Stationary Gaussian Process Example*: Comparison between the performance of REDS and GP regression (kriging) on the simulated data in Figure 1 that is generated from a the one-dimensional stationary GP. The accuracy metrics (RMSE – root mean square error; MAE – mean absolute error; CO – coverage; CRPS – continuous rank probability score; IS − interval score) demonstrate the prediction power of the REDS method is reasonable compared to the performance of the kriging model with known covariance function, although REDS attain slightly higher interval score.

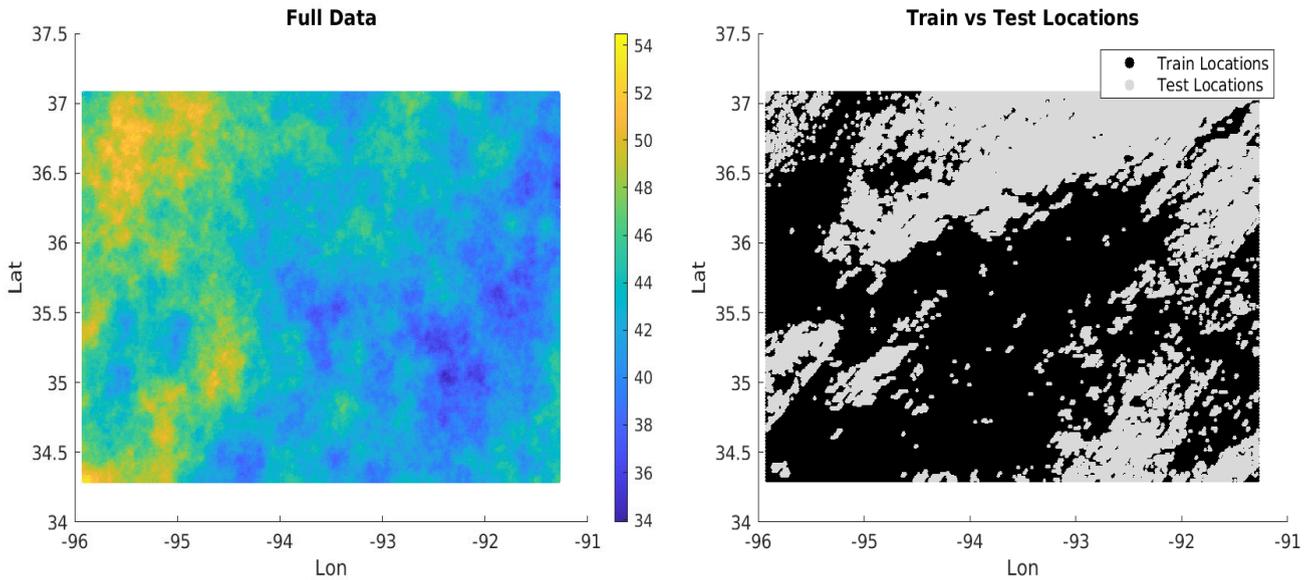

**Figure 2** **High Volume Simulated Two-Dimensional Data from a GP**: The left panel shows the full simulated data from Heaton et al. (2019). The right panel shows the training data locations (in black) and test data locations (in grey) used to fit the spatial models.

## 4.2 | Simulation: High Volume Two-Dimensional Gaussian Process

To demonstrate REDS effectiveness on a high volume spatial dataset, we consider the simulated data discussed in Heaton et al. (2019). These data were simulated using parameters that were estimated from a randomly selected subset from a true satellite image of temperature, which we use in the real-world example in Section 4.3. The simulated data are on a regular grid of size $500 \times 300$. The missing locations are kept the same as in Heaton et al. (2019) to facilitate comparison to the results presented therein. Specifically, Figure 2 shows the distribution of the $105,569$ training locations and $44,431$ test locations.

To apply REDS, we reformat the data to lie within a square grid with domain $[0, 1]^2$. We compare our results to the state-of-the-art methods that were applied to the same dataset in Heaton et al. (2019). The summary metrics from REDS are within the range of those presented for the state-of-the-art methods in Table 2. Similarly, the prediction and uncertainty maps shown in Figures 3 and 4 suggest that the model is providing reasonable predictions. However, like most basis function approaches, the predictions shown in Figure 4 have somewhat smoothed over the very fine-scale structures, but the uncertainty bounds reasonably capture the associated uncertainty.

## 4.3 | Real World Satellite Data Prediction: Application to a Large Spatial Dataset

Following the example in Section 4.2 that was used for comparing spatial prediction methodologies, we consider the true satellite data corresponding to land surface temperatures presented in Heaton et al. (2019). The data were obtained by the Terra instrument

| Metric | MAE | RMSE | Coverage | CRPS | IS |
|---|---|---|---|---|---|
| REDS | 0.8860 | 1.1232 | 0.9626 | 0.6356 | 5.6257 |
| Other methods in Heaton et al. (2019) | $0.61 - 1.06$ | $0.83 - 1.43$ | $0.44 - 1.00$ (0.96 best) | $0.43 - 0.76$ | $3.64 - 18.01$ |

**Table 2** *Comparisons of Predictive Metrics for High Volume Simulated Data from a Two-Dimensional GP:* Results from the application of REDS on the simulated data at Figure 2 are shown here. The first row shows the prediction summaries (RMSE – root mean square error; MAE – mean absolute error; CO – coverage; CRPS – continuous rank probability score; IS – interval score) from the REDS procedure and the bottom row shows the range of corresponding summary values obtained by other competing algorithms from that paper.

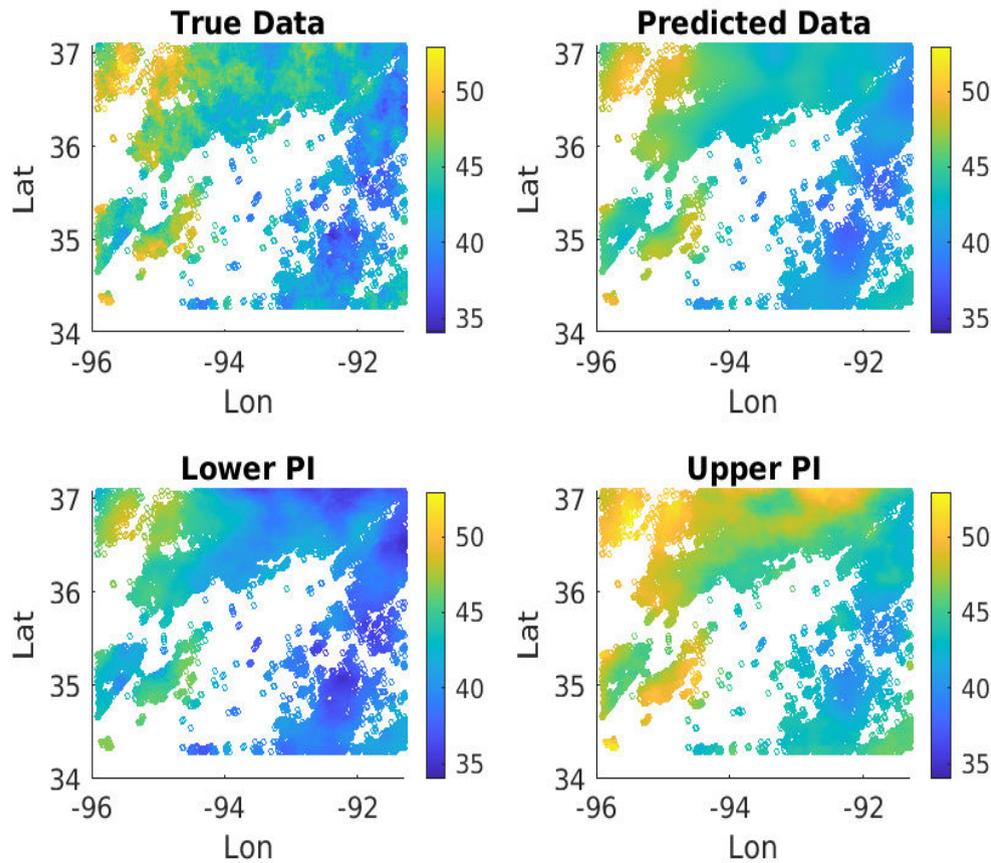

**Figure 3** **High Volume Simulated Data and Predictions from a Two-Dimensional GP**: The top left panel shows the true simulated values from Heaton et al. (2019) at the test locations. The top right panel shows the predicted values at these locations from REDS. The bottom row presents the uncertainty bounds associated with the REDS predictions; specifically, the left bottom and right bottom images correspond to the point-wise lower 2.5%-tiles and upper 97.5%-tiles, respectively.

on board the MODIS satellite from August 4, 2016 (Level-3 data) and is available for download using the MODIS re-projection tool web interface (MRTweb)[1]. The collected observations are given on an equidistant $500 \times 300$ rectangular grid between longitude $-95.91153$ to $-91.28381$ and latitude 34.29519 to 37.06811, out of which $105,569$ observations were used as training data. Note that the difference between this dataset and the simulated data in Section 4.2 is that the simulated data were generated following an assumed stationary Gaussian process, in which the mean surface was subtracted from the true data to make the data stationary in space. Importantly, REDS does not require that we subtract a linear effect of locations since it can automatically

---

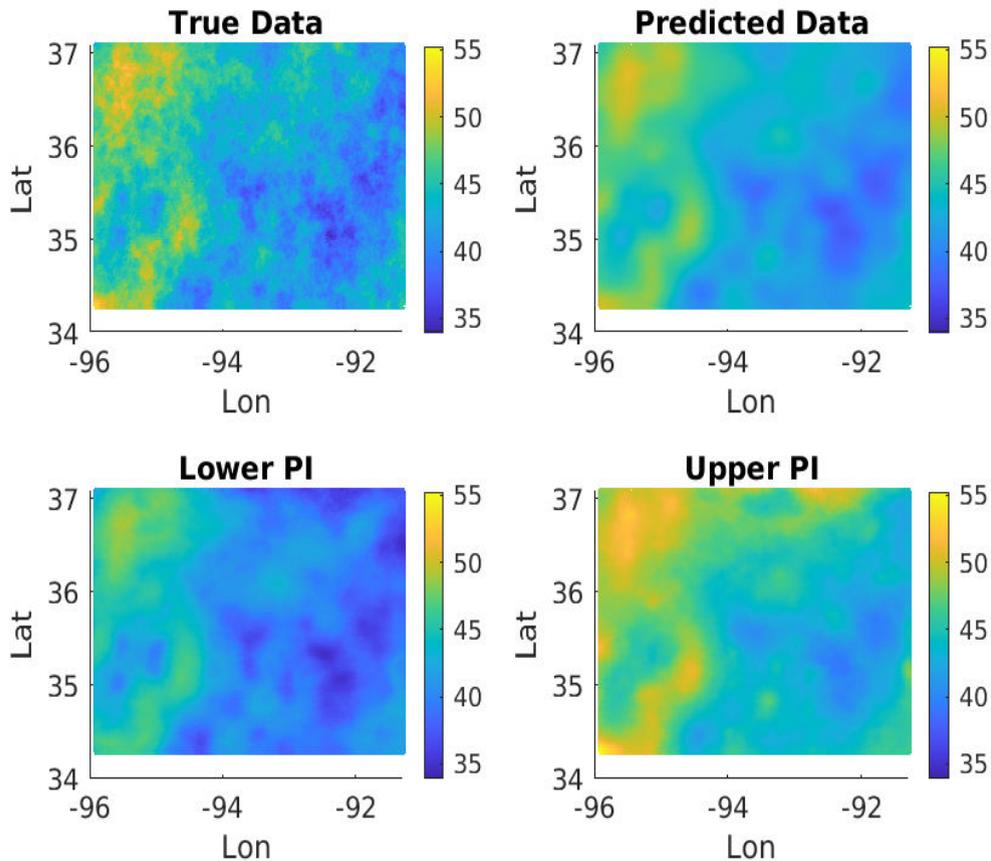

**Figure 4 High Volume Simulated Data and Predictions from a GP**: The top left panel shows the true simulated data from Heaton et al. (2019) at all locations. The top right panel shows the predicted surface from REDS. The bottom row shows the uncertainty bounds associated with the REDS prediction; specifically, the left bottom and right bottom images correspond to the point-wise lower 2.5%-tiles and upper 97.5%-tiles, respectively.

pick the necessary features from the over-complete basis formulation. Nevertheless, for demonstration, we also consider the case where a linear effect in terms of latitude and longitude is first removed from the data, the residuals are then considered as data in REDS with associated predictions, and the linear effect is added back.

Table 3 shows that REDS provides RMSE, MAE, coverage, and interval scores that are within the range of values from the state-of-the-art models in Heaton et al. (2019). In addition, note that the REDS results for the case where the mean is removed first are essentially the same as the case when we do not remove the mean first, suggesting that REDS is robust to stationarity in this case. Figure 5 shows the prediction surfaces and prediction intervals obtained from REDS for the missing (hold out) observations. The spatial plots are in agreement with the metrics in the sense that the predictions are visually similar to the hold out sample, and the prediction intervals show reasonable coverage. As with the simulated case in Section 4.2, the REDS procedure does appear to filter out very small scale spatial features. These results provide strong support that randomization methods such as REDS are able to learn how to predict high volume spatial data using a completely random initialization mechanisms and perform in the range of state-of-the-art spatial prediction methods.

## 5 | DISCUSSION

We develop a spatial prediction methodology by using random feature expansions in a hierarchical (or deep) manner, and then select the necessary features and corresponding expansion coefficients by employing a penalized regression (lasso). The methodology is applicable to continuous-valued spatial data, although it can easily be extended to other data types by changing the regression model in the last layer. The proposed methodology uses a rejection ensemble mechanism where for every ensemble member, the input data are first converted to a random basis expansion layer using a random Fourier transformation. Then, these transformations are passed through several random projection layers as with extreme learning machines. The final stage is trained using a penalized lasso regression with minibatch samples. This is performed for each ensemble member and the training RMSE is computed for each. The poorly performing random networks with the highest RMSEs are then rejected and the remaining networks are used to obtain the predictions as the point-wise medians. Uncertainty quantification is implemented by computing the IQR of the point predictions from these remaining "good" networks. Then, an optimal cutoff constant is calibrated such that the training data coverage matches a target Type I error level. It is important to note that other than the last layer in which a penalized regression is used, and the calibration of the uncertainty quantification, there is no learning in this approach – the basis function and deep layer weight matrices are all generated randomly.

| Metric | MAE | RMSE | Coverage | CRPS | IS |
|---|---|---|---|---|---|
| REDS (No mean adjustment) | 2.0105 | 2.3540 | 0.9341 | 1.3673 | 11.1448 |
| REDS (Mean adjustment) | 2.0213 | 2.3687 | 0.9295 | 1.3747 | 11.3419 |
| Other methods in Heaton et al. (2019) | $1.10 - 2.08$ | $1.53 - 2.64$ | $0.36 - 0.97$ (0.95 best) | $0.83 - 1.55$ | $7.44 - 34.78$ |

**Table 3** *Prediction Results for Satellite Temperature Data*: The top row shows REDS prediction and uncertainty quantification summary metrics (RMSE – root mean square error; MAE – mean absolute error; CO – coverage; CRPS – continuous rank probability score; IS – interval score) from prediction on the high volume satellite temperature dataset in Heaton et al. (2019). The second row shows the REDS prediction when a linear effect of latitude and longitude is removed first (see text). The bottom row presents the range of corresponding values obtained by other competing algorithms from the Heaton et al. (2019) paper.

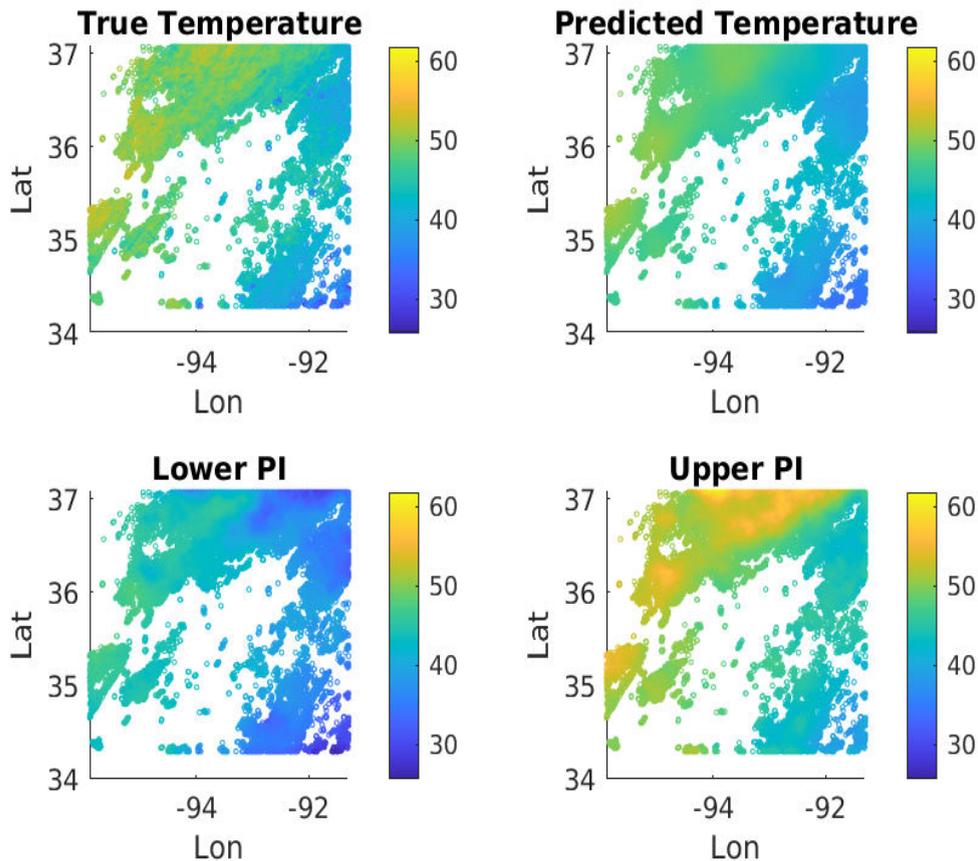

**Figure 5** **Prediction Images for Satellite Temperature Data Temperature**: REDS is applied to the large (500×300 grid) satellite temperature dataset from Heaton et al. (2019). The top left panel shows the observed temperature at the missing locations, which are the same as in the simulated example from section 4.2. The top right panel shows the plot of the predicted temperatures from REDS at the missing data locations. The bottom panels shows the associated REDS prediction intervals (i.e., the left bottom and right bottom images correspond to the point-wise lower 2.5%-tiles and upper 97.5%-tiles, respectively).

As with all "new" algorithms in machine learning, REDS is motivated by other models and so its structure is similar to many existing algorithms in statistics and machine learning. First, REDS can thought of as a fully randomized version of DeepKriging Chen et al. (2020) that combines basis transformations and deep learning to provide predictions for spatial data. The primary difference with REDS is that the deep learning comes from the random projection layers (extreme learning machines), and the basis transformation is based on flexible RFFs. Indeed, due to the presence of the RFF layer, REDS is directly related to other basis transformation based spatial models where prediction is a linear function of a deterministic family of basis transformations (e.g., fixed rank kriging, Cressie & Johannesson 2008). In addition, REDS has a connection to the so-called random kitchen sink algorithm (Rahimi & Recht 2008). The random kitchen sink prediction is a linear function of RFFs, whereas the REDS prediction further transforms the RFFs through the deep random projection layers to add flexibility. There are also similarities to (deep) kernel extreme learning machines (Huang, Zhou, Ding, & Zhang 2011) in the sense that REDS uses a RFF instead of a fixed transformation in the first layer before considering deep hidden random projections layers.

The purpose of the REDS algorithm is to provide a cost-effective and scalable algorithm that requires minimal training and yet can represent complex high-volume dependent data. Like all deep learning models, the algorithm requires the specification of some tuning parameters – but, importantly, the ensemble approach allows us to simulate over these if desired. REDS is a greedy rejection algorithm in the sense that we reject low-performing networks by studying the training RMSE after performing simple regularized model fitting on small minibatch samples of the data. Furthermore, the ensemble predictions and uncertainty quantification are based on robust statistics (median and IQR) to protect against poor networks. With all these carefully chosen steps, the algorithm performs well in terms of both accuracy and UQ. Indeed, its success is demonstrated by providing spatial predictions (point and interval coverage) that are within the range of comparison metrics to many state-of-the-art algorithms when compared with simulated datasets and a large satellite based image dataset. Average computation time for REDS is between $5 - 6$ minutes for the large dataset from Heaton et al. (2019), which is among the best candidates presented in the paper.

The REDS approach represents a simple attempt to perform random deep learning for spatial data. Despite its promise, there are several areas that can be improved and warrant future research. For example, although we have considered spatial data here, the method can easily be used for temporal data, spatio-temporal data, and other high-dimensional domains. Furthermore, different data types can also be considered by substituting the regression training in the final layer of the network with other penalized loss training. Plausible hyper-parameters were chosen by studying various simulation runs to build intuition on model performance and then fixing some and letting the most important ones be sampled from subjectively-specified distributions for each ensemble member. It is likely that this procedure can be made more systematic with additional research.

In addition, our ensemble based prediction intervals are often non-smooth, and optimal approaches to smooth them are the subject of future research. Also, although the model is computationally less costly than many spatial prediction algorithms, performing a penalized regression for each ensemble still requires moderate computational cost for large datasets, which likely can be improved by considering scalable regression algorithms such as greedy variable selection. Finally, from a theory perspective, it will be important to more formally consider the choices of hyper-parameters and the convergence rates for specific families of functions.

# ACKNOWLEDGMENTS

This research was partially supported by the U.S. National Science Foundation (NSF) grant SES-1853096. The computation for this work was performed on the high performance computing infrastructure provided by the Research Computing Support Services at the University of Missouri, Columbia, MO, and is supported in part by the NSF grant CNS-1429294.

## Author contributions

RD conceptualized the methodology, implemented it, and drafted the manuscript. CKW contributed to discussions of the methodology and helped polish the manuscript.

## Financial disclosure

None reported.

## Conflict of interest

The authors declare no potential conflict of interests.

## AUTHOR BIOGRAPHY

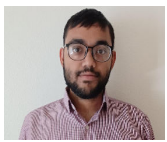

**Ranadeep Daw** is a Ph.D. candidate in the Department of Statistics at the University of Missouri.

**Christopher K. Wikle** is Curators' Distinguished Professor of Statistics and Department Chair at the University of Missouri.